\documentclass[icps3]{svjour}
\usepackage{epsfig}
\usepackage{times,mathptm}  %  use Times fonts, not LaTeX

\begin{document}
\title{Critical exponents for the metal-insulator transition 
of $^{70}$Ge:Ga in magnetic fields}
\titlerunning{Critical exponents for the metal-insulator transition 
of Ge:Ga in magnetic fields}
% The running title should be in less than 70 characters.
%
\author{
Michio Watanabe\inst{1}\thanks{(post-doctoral) JSPS Research Fellow}, 
Kohei M. Itoh\inst{1}\thanks{also at PRESTO-JST}, 
Masashi Morishita\inst{2}, 
Youiti Ootuka\inst{2}, 
Eugene E. Haller\inst{3}
}                     % Do not remove
%
% Insert author list for the running head here
\authorrunning{Michio Watanabe et al.}
% If the number of the authors is more than 3, only the first author
% should be listed and the others represented as et al.
%
\institute{
Dept. Applied Physics and Physico-Informatics, Keio University, 
3-14-1 Hiyoshi, Kohoku-ku, Yokohama 223-8522, Japan 
\and 
Institute of Physics, University of Tsukuba, 1-1-1 Tennodai, 
Tsukuba, Ibaraki 305-8571, Japan
\and 
Lawrence Berkeley National Laboratory and University of California 
at Berkeley, Berkeley, California 94720, USA}
\maketitle
\begin{abstract}
We have measured the electrical conductivity of nominally 
uncompensated $^{70}$Ge:Ga samples in magnetic fields up to 
$B=8$~T at low temperatures ($T=0.05-0.5$~K) in order to 
investigate the metal-insulator transition in magnetic fields.  
The values of the critical exponents in magnetic fields 
are consistent with the scaling theories.  
\end{abstract}
\section{Introduction}
Doped crystalline semiconductors are ideal solids to probe 
the effects of both disorder and electron-electron interaction 
on the metal-insulator transition (MIT) in disordered electronic 
systems~\cite{Bel94}.  Important information about the MIT is provided 
by the critical exponent $\mu$ for the zero-temperature conductivity 
$\sigma(0)$ defined by 
\begin{equation}
\sigma(0) \propto (N-N_c)^\mu, 
\end{equation}
where $N$ is the dopant concentration and $N_c$ is the critical 
concentration for the MIT, and $\mu\approx0.5$ has been found in a 
number of nominally uncompensated semiconductors including 
our $^{70}$Ge:Ga~\cite{Ito96,Wat98}.  

According to the theories of the MIT, the value of the critical exponents 
does not depend on the details of the system, but depends only on the 
universality class to which the system belongs.  
In this sense, the application of a magnetic field is important 
because the motion of carriers loses its time-reversal symmetry 
in magnetic fields, and the universality class changes.  
In our earlier work, 
we reported that a different exponent $\mu=\mu'=1.1\pm0.1$ 
is obtained in magnetic fields~\cite{Wat99}.  
Here, $\mu'$ characterizes the magnetic-field-induced MIT: 
\begin{equation}
\sigma(0)\propto(B_c-B)^{\mu'}.  
\label{eq:Bmit}
\end{equation}
Since this result is based solely on the metallic samples, 
in this work we perform a finite-temperature scaling 
analysis~\cite{Bel94} which uses the data on the both sides 
of the transition.  
Moreover, we investigate the temperature dependence 
of the conductivity on the insulating side 
in the context of variable-range-hopping (VRH) conduction.   
\section{Experiment}
All the samples were prepared by neutron-transmutation-doping 
(NTD) of isotopically enriched $^{70}$Ge single crystals.  
The NTD method assures a homogeneous Ga acceptor distribution 
which is a crucial condition for experimental studies 
of the MIT~\cite{Ito96,Wat98}.  
The electrical conductivity was measured at low temperatures 
between 0.05 and 0.5~K.  
\section{Results and discussion}
\begin{figure}
\begin{center}
\epsfig{file=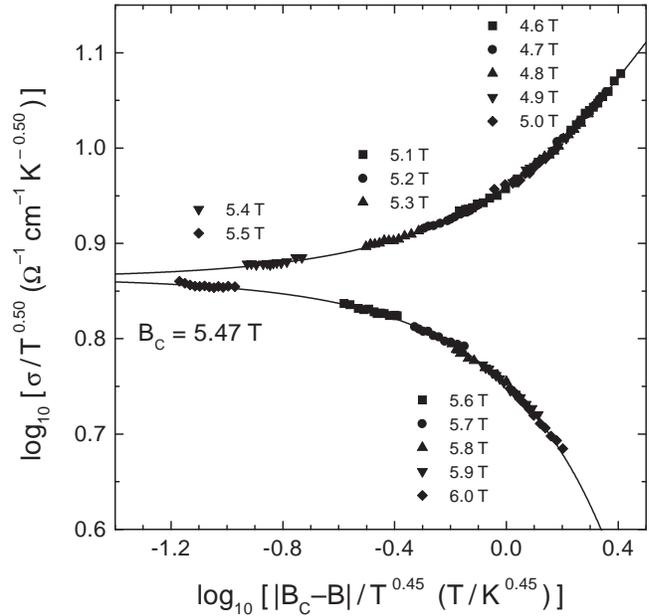,width=\columnwidth,
bbllx=37,
bblly=283,
bburx=507,
bbury=737,clip=,
angle=0}
\end{center}
\caption
{
Finite-temperature scaling plot 
for the magnetic-field-induced metal-insulator transition 
in the sample having $N=2.004\times10^{17}$~cm$^{-3}$.  
}
\label{fig:finiteT}
\end{figure}

We show in Fig.~\ref{fig:finiteT} that 
$\mu'=1.1$ (Ref.~\ref{Wat99}) yields an 
excellent finite-temperature scaling plot 
\begin{equation}
\frac{\sigma(B,T)}{T^x}
\propto f\left(\frac{|B_c-B|}{T^y}\right), 
\end{equation} 
where $x/y$ is equivalent to $\mu'$.  
%
%%%%%%%%%%%%%%%%%%%%%%%%%%%%%%%%%
%%%%%%  Fig. 1             %%%%%%
%%%%%%%%%%%%%%%%%%%%%%%%%%%%%%%%%
%%%%%%%%%%%%%%%%%%%%%%%%%%%%%%%%%
%
Here we employ $B_c$ obtained by fitting Eq.~(\ref{eq:Bmit}).  
The temperature variation of the conductivity is proportional 
to $T^{1/2}$ even around the critical point in magnetic 
fields~\cite{Wat99}, leading to $x=1/2$.  
Note that $y=x/\mu'=0.45$, i.e., none of these parameters 
are used as a fitting parameter.  
Hence, Fig.~\ref{fig:finiteT} 
strongly supports $\mu'=1.1$.  

\begin{figure}
\begin{center}
\epsfig{file=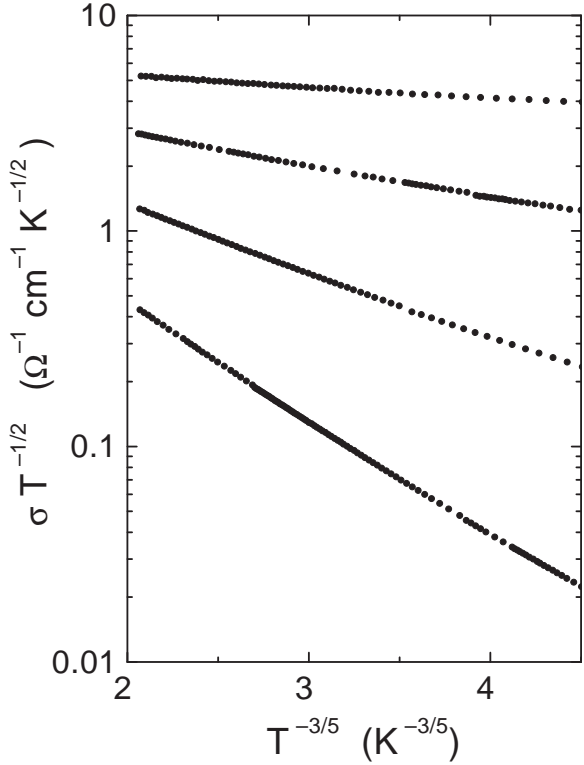,width=0.9\columnwidth,
bbllx=28,
bblly=59,
bburx=538,
bbury=733,clip=,
angle=0}
\end{center}
\caption
{
Conductivity multiplied by $T^{-1/2}$ as a function of 
$T^{-3/5}$ for the sample having $N=1.912\times10^{17}$~cm$^{-3}$ 
in magnetic fields.  From top to bottom in units of tesla, 
the magnetic induction is 5, 6, 7, and 8, respectively.    
}
\label{fig:Tdep}
\end{figure}

\begin{figure}
\begin{center}
\epsfig{file=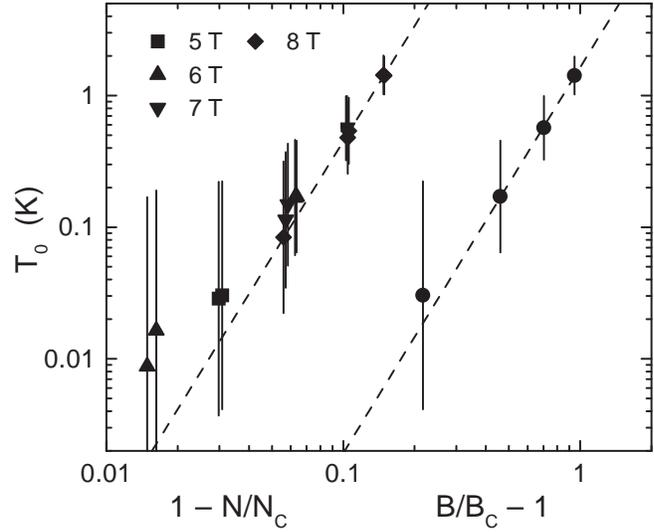,width=\columnwidth,
bbllx=24,
bblly=268,
bburx=558,
bbury=711,clip=,
angle=0}
\end{center}
\caption
{
$T_0$ determined by $\sigma(T)\propto 
T^{1/2}\exp[-(T_0/T)^{3/5}]$ 
as a function of $1-N/N_c(B)$ in constant 
magnetic fields of $B=5$, 6, 7, and 8~T 
(left data set), and as a function of $B/B_c-1$ for 
the sample having $N=1.912\times10^{17}$~cm$^{-3}$
(right data set).  The dashed lines represent 
the best fits to the data satisfying $T_0>0.05$~K.  
}
\label{fig:T0}
\end{figure}

The temperature dependence of the conductivity 
on the insulating side of the MIT is shown 
in Fig.~\ref{fig:Tdep}.  
%
%%%%%%%%%%%%%%%%%%%%%%%%%%%%%%%%%
%%%%%%  Fig. 2             %%%%%%
%%%%%%%%%%%%%%%%%%%%%%%%%%%%%%%%%
%%%%%%%%%%%%%%%%%%%%%%%%%%%%%%%%%
%
We already reported that VRH conductivity at $B=0$ obeys 
Efros and Shklovskii's (ES) law~\cite{Shk84} 
\begin{equation}
\sigma(N,0,T)=\sigma_0(N,0)\exp[-(T_0/T)^{1/2}] 
\end{equation}
even in the immediate vicinity of $N_c$ ($0.99N_c<N<N_c$) 
when an appropriate temperature dependence of the prefactor 
$\sigma_0\propto T^{r}$ is taken into account~\cite{Wat00}.  
Based on this finding, we analyze in this work the data 
at $B\geq5$~T in the context of ES VRH conduction 
\begin{equation}
\sigma(N,B,T)=\sigma_0(N,B)\exp[-(T_0/T)^{3/5}] 
\label{eq:VRHinB}
\end{equation}
in a strong field ($\sqrt{\hbar/eB}\ll\xi$, 
where $\xi$ is the localization length)~\cite{Shk84}.  
By assuming $\sigma_0\propto T^{1/2}$, which is consistent 
with the above finite-temperature scaling analysis, 
the conductivity in magnetic fields is described well 
by Eq.~(\ref{eq:VRHinB}) as seen in Fig.~\ref{fig:Tdep}.  
The values of $T_0$ in Eq.~(\ref{eq:VRHinB}) satisfy 
the relations $T_0\propto(N_c-N)^{\alpha}$ and 
$T_0\propto(B-B_c)^{\alpha'}$, and $\alpha=\alpha' 
\approx 2.9$ is obtained from the data satisfying 
$T_0>0.05$~K.  (See Fig.~\ref{fig:T0}.) 
%
%%%%%%%%%%%%%%%%%%%%%%%%%%%%%%%%%
%%%%%%  Fig. 3             %%%%%%
%%%%%%%%%%%%%%%%%%%%%%%%%%%%%%%%%
%%%%%%%%%%%%%%%%%%%%%%%%%%%%%%%%%
%
Note that the condition $T<T_0$ is required for the 
ES theory to be valid, i.e., $T_0$ has to be evaluated 
only from the data obtained at temperatures low enough 
to satisfy this condition.   
Since $T_0\propto(\xi\epsilon)^{-1}$ and both $\xi$ 
and the dielectric constant $\epsilon$ diverge at $N_c$, 
$\alpha=\nu+\zeta$.  Here, $\nu$ and $\zeta$ 
are given by $\xi\propto(N_c-N)^{-\nu}$ 
and $\epsilon\propto(N_c-N)^{-\zeta}$, respectively.  
The relation $2\nu\approx\zeta$, which was predicted 
theoretically~\cite{Kaw84}, has been obtained at $B=0$ 
for the present system by measuring magnetoresistance 
in weak fields ($\sqrt{\hbar/eB}\gg\xi$)~\cite{Wat00}.  
Assuming the relation $2\nu=\zeta$, 
$\nu\approx1$ is obtained for $B\neq0$, and hence, 
the Wegner relation $\mu=\nu$~\cite{Weg76} holds 
in magnetic fields.  
\section{Conclusion}
The critical exponent of the zero-temperature conductivity 
for the metal-insulator transition of doped semiconductors 
in magnetic fields is confirmed to be close to unity.  
Other exponents in magnetic fields are also explained 
within the context of the scaling theories.  
\section*{Acknowledgments}
We are thankful to T. Ohtsuki for valuable discussions.  
Conductivity measurements were carried out 
at the Cryogenic Center, the University of Tokyo, Japan.  
This work was supported by a Grant-in-Aid for Scientific Research 
from the Ministry of Education, Science, Sports, and Culture, Japan, 
the Director, Office of Energy Research, Office of 
Basic Energy Science, Materials Sciences Division of the U. S. 
Department of Energy under Contract No.~DE-AC03-76SF00098, 
and U. S. NSF Grant No.~DMR-97 32707.  
\end{document}